\documentclass[11pt,fleqn]{article}
\usepackage[top=2.5cm,bottom=2.5cm,left=2.5cm,right=2.5cm]{geometry}
\usepackage[T1]{fontenc}
\usepackage[ansinew]{inputenc}
\usepackage{lmodern}
\usepackage{graphicx}
\usepackage{amsmath}
\usepackage{amsthm}
\usepackage{amsfonts}
\usepackage{indentfirst}
\usepackage{amssymb}
\usepackage{setspace}
\usepackage{textcomp}
\usepackage{float}
\usepackage{url}

\newcommand{\ba}{\begin{array}}
\newcommand{\ea}{\end{array}}

\begin{document}
\newcommand{\be}{\begin{equation}}
\newcommand{\ee}{\end{equation}}
\newcommand{\bc}{\begin{center}}
\newcommand{\ec}{\end{center}}
\newcommand{\bdm}{\begin{displaymath}}
\newcommand{\edm}{\end{displaymath}}
\newcommand{\ds}{\displaystyle}
\newcommand{\p}{\partial}
\newcommand{\INT}{\int\limits}
\newcommand{\SUM}{\sum\limits}
\newcommand{\bfm}[1]{\mbox{\boldmath $ #1 $}}
\renewcommand{\theequation}{\arabic{section}.\arabic{equation}}

\title{ \bf Asymptotic analysis of drug dissolution in two layers having widely differing
diffusion coefficients
}
\author{
{\em  Michael Vynnycky$^{a}$, Sean McKee$^{b}$, Martin Meere$^{c}$, Chris McCormick$^{d}$, Sean McGinty$^{e}$}
\vspace{5mm}\\
$^{a}$Division of Processes,
Department of Materials Science and Technology,\\  KTH Royal Institute of
Technology,\\ Brinellv\"{a}gen 23, 100 44 Stockholm,\\ Sweden.
E-mail: {\tt michaelv@kth.se}\\
\vspace{1mm}\\
$^{b}$Department of Mathematics and Statistics,\\
University of Strathclyde,\\Glasgow, G1 1XH, UK. \\
\vspace{1mm}\\
$^{c}$Department of Applied Mathematics,\\
NUI Galway, Galway, Ireland.  \\
\vspace{1mm}\\
$^{d}$Department of Biomedical Engineering,\\
University of Strathclyde,\\ Glasgow, G4 0NW, UK. \\
\vspace{1mm}\\
$^{e}$Division of Biomedical Engineeering,\\
University of Glasgow, Glasgow, G12 8QQ, UK
}

\date{}

\maketitle





\begin{abstract}This paper is concerned with a diffusion-controlled moving-boundary problem
in drug dissolution, in which the moving front passes from one medium to
another for which the diffusion coefficient is many orders of magnitude
smaller. It has been shown in an earlier paper that a similarity solution
exists while the front is passing through the first layer, but that this
breaks down in the second layer. Asymptotic methods are used to understand
what is happening in the second layer. Although this necessitates numerical
computation, one interesting outcome is that only one calculation is required,
no matter what the diffusion coefficient is for the second layer.
\end{abstract}



\section{Introduction}

Moving boundary problems arise in many industrial applications and, as a
result, they have been studied extensively in the mathematical literature
(\cite{DPDM,HH,LM}). When the problem is well characterised by a
one-dimensional system of equations, analytical solutions are often readily
obtained. For example, if the system comprises a one-dimensional diffusion
equation with appropriate initial and boundary conditions, as well as a Stefan
condition to track the position of the moving boundary, then it can often be
shown that the problem is self-similar, and through a similarity reduction one
may convert the original system to a system of ordinary differential
equations. Some discussion of the analytical solution of moving boundary
problems arising in diffusive systems can be found in \cite{Crank}.

However, it is not always the case that such a similarity structure exists for
all time and often one has to resort to seeking a numerical solution using an
appropriate numerical method: for example, a front-tracking finite difference
scheme (\cite{Crank}). In this context, a recent development, which is
exploited in this work, is to analyze the governing partial differential
equations for small time, determine if there is a similarity solution and, if
there is, use it as an initial condition for the subsequent computation, which
is performed in terms of the similarity variables, rather than the original
physical variables; in particular, this approach is of importance for
maintaining the accuracy of a numerical scheme in problems where the initial
thickness of the domain of interest is zero (\cite{P38,P71,P88}), as will be
the case in this work.

Whilst a common type of moving boundary problem often involves phase change,
as in \cite{P38,P71,P88}, an arguably less common type is where there is no
phase change involved, but the front in question passes from one medium into
another; in this situation also, there can be no hope of a similarity solution
that is valid for all time. An example of an application where precisely this
problem arises is presented in the recent publication by \cite{Vo:2018}; a
particular characteristic of this problem is that the diffusion coefficient of
the second medium is several orders of magnitude smaller than that of the first.

\begin{figure}[ptb]
\centering\scalebox{0.5}{\includegraphics{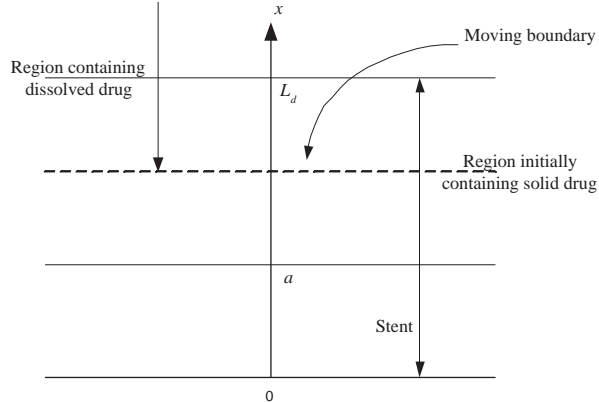}}
\caption{Schematic showing the problem considered by \cite{Vo:2018}. The
region $0<x<L_{d}$ initially contains drug at uniform concentration $c_{0}$.
For $t>0$, drug dissolves on a moving front (where the concentration is
identically $c_{s}$, the solubility of the drug), starting at $x=L_{d}$. Drug
dissolution is complete when the moving boundary tracks back to $x=0$.
Dissolved drug diffuses out of the system into a release medium which is
considered to be infinite. The diffusion coefficient of the dissolved drug in
the region $0<x<a$ is much smaller than that in the region $x>a$. }%
\label{fig0}%
\end{figure}

\cite{Vo:2018} investigated the drug release from polymer-free coronary stents
with microporous surfaces. The investigation was both experimental and
theoretical. As part of the theoretical analysis, the following
one-dimensional diffusion problem arose:%
\begin{equation}
\frac{\partial c}{\partial t}=\frac{\partial}{\partial x}\left(  D\left(
x\right)  \frac{\partial c}{\partial x}\right)  ,\qquad x>s\left(  t\right)
,\text{ }t>0, \label{eq1}%
\end{equation}%
\begin{align}
c  &  =c_{s},\quad-D\left(  x\right)  \frac{\partial c}{\partial x}%
=\frac{\mathrm{d}s}{\mathrm{d}t}(c_{s}-c_{0})\quad\text{at }x=s\left(
t\right)  ,t>0,\label{eq2}\\
c  &  \rightarrow0\quad\text{as \ }x\rightarrow\infty,\text{ }t>0, \label{eq3}%
\end{align}%
\begin{equation}
s\left(  0\right)  =L_{d},\quad c\left(  x,0\right)  =0\quad\text{for }%
x>L_{d}. \label{eq4}%
\end{equation}
Here, $c$ represents the concentration of the drug, $s\left(  t\right)  $ a
free surface between the dissolved and undissolved drug, $L_{d}$ denotes the
thickness of the drug layer initially, which occupies the region $0<x<L_{d}$,
$a<L_{d}$ denotes the mean position of the microporous region (also containing
drug), $c_{s}$ the solubility of the drug and $c_{0}$ the initial constant
concentration for $x<L_{d}.$ The spatially dependent diffusion coefficient is%
\begin{equation}
D\left(  x\right)  =\left\{
\begin{array}
[c]{ll}%
D_{e}\left(  <D_{w}\right)  & \text{if }0<x\leq a_{-}\\
D_{w} & \text{if }x\geq a_{+}%
\end{array}
\right.  . \label{eq5}%
\end{equation}

The problem given by (\ref{eq1}-\ref{eq5}) gives rise to a two-stage release
of drug (Figure 1). In Stage 1, the drug dissolves on a moving front in the
region $a<x<L_{d}$ and diffuses out of the system. In Stage 2, the moving
boundary has tracked back to $x=a$ and the drug then proceeds to dissolve from
the rough surface region where it is released at a slower rate. For Stage 1
($s\left(  t\right)  >a$), \cite{Vo:2018} wrote down an analytical solution,
the derivation of which may be found in \cite{McGinty:2015}. The solution is
given by%
\begin{equation}
s\left(  t\right)  =L_{d}-\theta\sqrt{t},\quad c\left(  x,t\right)
=\frac{c_{s}\text{erfc}\left(  \frac{x-L_{d}}{2\sqrt{D_{w}t}}\right)
}{\text{erfc}\left(  -\frac{\theta}{2\sqrt{D_{w}}}\right)  },\quad
L_{d}-\theta\sqrt{t}<x<\infty,\text{ \ }0<t<t_{a}, \label{1-6}%
\end{equation}
\ \ where $\theta$ is determined by%
\begin{equation}
\frac{\theta}{2\sqrt{D_{w}}}\exp\left(  \frac{\theta^{2}}{4D_{w}}\right)
\text{erfc}\left(  -\frac{\theta}{2\sqrt{D_{w}}}\right)  =\frac{1}{\sqrt{\pi}%
}\frac{c_{s}}{c_{0}-c_{s}}. \label{1-7}%
\end{equation}
The solution is valid until $t=t_{a},$ whereupon $s\left(  t_{a}\right)  =a,$
so that
\begin{equation}
t_{a}=\frac{\left(  L_{d}-a\right)  ^{2}}{\theta^{2}}. \label{theta}%
\end{equation}
Furthermore, at $t=t_{a},$%
\begin{equation}
c\left(  x,t_{a}\right)  =c_{a}\left(  x\right)  =\frac{c_{s}\text{erfc}%
\left(  \frac{x-L_{d}}{2\sqrt{D_{w}t_{a}}}\right)  }{\text{erfc}\left(
-\frac{\theta}{2\sqrt{D_{w}}}\right)  },\quad a\leq x<\infty. \label{1-9}%
\end{equation}
For Stage 2, a numerical procedure was employed.

In this paper, we will be concerned with the release of drug from the system
during Stage 2. In particular, we adopt an asymptotic approach to derive
approximate solutions for this phase of release. In Section \ref{sec2}, we
start by presenting the equations that represent Stage 2 of the release. We
then outline our asymptotic argument. In Section \ref{sec3}, we provide
results including comparisons with the numerical solutions obtained by
\cite{Vo:2018}.

\section{Stage 2 $\left(  s\left(  t\right)  <a\right)  \label{sec2}$}

The Stage 2 problem when $t>t_{a}$ may then be formulated in dimensional form
as:%
\begin{align}
\frac{\partial c}{\partial t}  &  =\frac{\partial}{\partial x}\left(
D_{w}\frac{\partial c}{\partial x}\right)  ,\qquad a<x<\infty,\quad
t>t_{a},\label{a1}\\
\frac{\partial c}{\partial t}  &  =\frac{\partial}{\partial x}\left(
D_{e}\frac{\partial c}{\partial x}\right)  ,\qquad s(t)<x<a,\quad
t>t_{a},\label{a2}\\
c  &  =c_{s},\quad-D_{e}\frac{\partial c}{\partial x}=\frac{\mathrm{d}%
s}{\mathrm{d}t}(c_{s}-c_{0}),\quad\text{at }x=s\left(  t\right)  ,\label{a3}\\
c  &  \rightarrow0,\quad\text{as }x\rightarrow\infty,\label{a4}\\
s\left(  t_{a}\right)   &  =a,\quad c\left(  x,t_{a}\right)  =c_{a}\left(
x\right)  ,\quad x\geq a. \label{a5}%
\end{align}
In addition, we require%
\begin{align}
\left[  c\right]  _{-}^{+}  &  =0\quad\text{at }x=a,\label{a6}\\
\left(  D_{e}\frac{\partial c}{\partial x}\right)  _{-}  &  =\left(
D_{w}\frac{\partial c}{\partial x}\right)  _{+}\quad\text{at }x=a. \label{a7}%
\end{align}

We non-dimensionalize the problem by setting
\begin{equation}
X=\frac{x}{a},\quad T=\frac{t-t_{a}}{a^{2}/D_{e}},\quad S=\frac{s}{a},\quad
C=\frac{c}{c_{s}},\quad C_{a}=\frac{c_{a}}{c_{s}}. \label{three-eight}%
\end{equation}
This gives
\begin{align}
\delta\frac{\partial C}{\partial T}  &  =\frac{\partial^{2}C}{\partial X^{2}%
},\qquad1<X<\infty,\quad T>0,\label{a8}\\
\frac{\partial C}{\partial T}  &  =\frac{\partial^{2}C}{\partial X^{2}},\qquad
S(T)<X<1,\quad T>0,\label{a9}\\
C  &  =1,\quad-\frac{\partial C}{\partial X}=\frac{\mathrm{d}S}{\mathrm{d}%
T}(1-\frac{c_{0}}{c_{s}}),\quad\text{at }X=S\left(  T\right)  ,\label{a10}\\
C  &  \rightarrow0,\quad\text{as }X\rightarrow\infty,\label{a11}\\
S\left(  0\right)   &  =1,\quad C\left(  X,0\right)  =C_{a}\left(  X\right)
,\quad X\geq1, \label{a12}%
\end{align}
where $\delta=D_{e}/D_{w}\ll1,$ as in \cite{Vo:2018}, and%
\begin{equation}
C_{a}\left(  X\right)  =\frac{\text{erfc}\left(  \frac{aX-L_{d}}{2\sqrt
{D_{w}t_{a}}}\right)  }{\text{erfc}\left(  -\frac{\theta}{2\sqrt{D_{w}}%
}\right)  }. \label{a12a}%
\end{equation}
In addition, we have%
\begin{align}
\left[  C\right]  _{-}^{+}  &  =0\quad\text{at }X=1,\label{a13}\\
\delta\left(  \frac{\partial C}{\partial X}\right)  _{-}  &  =\left(
\frac{\partial C}{\partial X}\right)  _{+}\quad\text{at }X=1. \label{a14}%
\end{align}

We have%
\begin{align}
\frac{\partial^{2}C}{\partial X^{2}}  &  \approx0,\quad0<X<\infty
,\label{a15}\\
C  &  \rightarrow0,\quad\text{as }X\rightarrow\infty,\label{a16}\\
\left(  \frac{\partial C}{\partial X}\right)  _{+}  &  \approx\;0\quad\text{at
}X=1, \label{a17}%
\end{align}
which would require $C\equiv0,$ for $X>1.$ For $X<1,$ we would have%
\begin{align}
\frac{\partial C}{\partial T}  &  =\frac{\partial^{2}C}{\partial X^{2}},\qquad
S(T)<X<1,\quad T>0,\label{a18}\\
C  &  =1,\quad-\frac{\partial C}{\partial X}=\frac{\mathrm{d}S}{\mathrm{d}%
T}(1-\frac{c_{0}}{c_{s}}),\quad\text{at }X=S\left(  T\right)  . \label{a19}%
\end{align}
Also, (\ref{a13}) would imply
\[
C=0\text{ at }X=1.
\]
In fact, this cannot hold for all time, since $C=1$ at $X=1$ at $T=0,$ i.e. in
dimensional form, $c=c_{s}$ when $x=s\left(  t_{a}\right)  =a.$


\subsection{Asymptotic argument}

The above suggests that we must try to retain the term on the left-hand side
of (\ref{a8}), which can be achieved if $T\sim\delta.$ This will mean that the
left-hand side of (\ref{a9}) will be large, and would need to be balanced by
the right-hand side, indicating that $1-X,$ i.e. the width of the lower
region, must be of an appropriately small width. Thus, we suppose that
$1-X\sim\left[  X\right]  ,$ where $\left[  X\right]  \ll1,$ and is still to
be determined. Thus, with
\begin{equation}
1-X=\left[  X\right]  \tilde{X},\quad1-S=\left[  X\right]  \tilde{S},\quad
T=\delta\tilde{T}, \label{te2}%
\end{equation}
we have%
\begin{align}
\frac{\partial C}{\partial\tilde{T}}  &  =\frac{\partial^{2}C}{\partial X^{2}%
},\qquad1<X<\infty,\quad\tilde{T}>0,\label{fr0}\\
\frac{\left[  X\right]  ^{2}}{\delta}\frac{\partial C}{\partial\tilde{T}}  &
=\frac{\partial^{2}C}{\partial\tilde{X}^{2}},\qquad\tilde{X}>0,\quad\tilde
{T}>0, \label{fr00}%
\end{align}
subject to%
\begin{align}
C  &  \rightarrow0\quad\text{as }X\rightarrow\infty,\\
C  &  =1,\quad-\frac{\partial C}{\partial\tilde{X}}=\frac{\left[  X\right]
^{2}}{\delta}\frac{\mathrm{d}\tilde{S}}{\mathrm{d}\tilde{T}}(1-\frac{c_{0}%
}{c_{s}}),\quad\text{at }\tilde{X}=\tilde{S}\left(  \tilde{T}\right)
,\label{fr000}\\
\tilde{S}  &  =0,\quad C=C_{a}\left(  X\right)  ,\quad\text{at }\tilde
{T}=0,\text{ }X>1\left(  \tilde{X}<0\right)  .
\end{align}
In addition, we have%
\begin{align}
\left[  C\right]  _{-}^{+}  &  =0\quad\text{at }X=1,\left(  \tilde{X}=0\right)
\\
-\frac{\delta}{\left[  X\right]  }\left(  \frac{\partial C}{\partial\tilde{X}%
}\right)  _{\tilde{X}=0}  &  =\left(  \frac{\partial C}{\partial X}\right)
_{X=1}. \label{fr}%
\end{align}

We must now choose $\left[  X\right]  $ so that (\ref{fr0})-(\ref{fr})
constitute a self-consistent system. There are basically only two
possibilities: $\left[  X\right]  \sim\delta$ and $\left[  X\right]
\sim\delta^{1/2}.$ We try these in turn.

\subsubsection{$\left[  X\right]  \sim\delta$}

Equation (\ref{fr00}) gives
\begin{equation}
\frac{\partial^{2}C}{\partial\tilde{X}^{2}}=0, \label{r0}%
\end{equation}
subject to, from (\ref{fr000}),
\begin{equation}
C=1,\quad\frac{\partial C}{\partial\tilde{X}}=0,\quad\text{at }\tilde
{X}=\tilde{S}\left(  \tilde{T}\right)  \label{r00}%
\end{equation}
and%
\begin{align}
\left[  C\right]  _{-}^{+}  &  =0\quad\text{at }X=1,\label{r1}\\
-\left(  \frac{\partial C}{\partial\tilde{X}}\right)  _{\tilde{X}=0}  &
=\left(  \frac{\partial C}{\partial X}\right)  _{X=1}. \label{r11}%
\end{align}
Thus, (\ref{r0}) and (\ref{r00}) give just $C\equiv1$ for $X<1,$ which means
that (\ref{r1})\ and (\ref{r11}) would become%
\begin{align}
C  &  =1\quad\text{at }X=1,\label{r2}\\
\left(  \frac{\partial C}{\partial X}\right)  _{X=1}  &  =0. \label{r22}%
\end{align}
Clearly what we have obtained is not self-consistent: $C$ for $X>1$ must
satisfy two boundary conditions, (\ref{r2}) and (\ref{r22}), at $X=1,$ which
is clearly not possible, and $\tilde{S}\left(  \tilde{T}\right)  $ remains undetermined.

\subsubsection{$\left[  X\right]  \sim\delta^{1/2}\label{pah}$}

With $\left[  X\right]  \sim\delta^{1/2},$ we have
\begin{equation}
\frac{\partial C}{\partial\tilde{T}}=\frac{\partial^{2}C}{\partial X^{2}%
},\qquad1<X<\infty,\quad\tilde{T}>0, \label{t}%
\end{equation}
subject to
\begin{equation}
C\rightarrow0,\quad\text{as }X\rightarrow\infty, \label{t0}%
\end{equation}
and, from (\ref{fr}),%
\begin{equation}
\frac{\partial C}{\partial X}=0\quad\text{at }X=1. \label{to}%
\end{equation}
Also, (\ref{fr00}) becomes
\begin{equation}
\frac{\partial C}{\partial\tilde{T}}=\frac{\partial^{2}C}{\partial\tilde
{X}^{2}},\qquad\tilde{X}>0,\quad\tilde{T}>0, \label{eq34a}%
\end{equation}
subject to%
\begin{align}
C  &  =C_{+}\left(  \tilde{T}\right)  \quad\text{at }\tilde{X}=0,
\label{eq34b}\\
C  &  =1,\quad-\frac{\partial C}{\partial\tilde{X}}=\frac{\mathrm{d}\tilde{S}%
}{\mathrm{d}\tilde{T}}(1-\frac{c_{0}}{c_{s}}),\quad\text{at }\tilde{X}%
=\tilde{S}\left(  \tilde{T}\right)  , \label{eq34v}%
\end{align}
where%
\begin{equation}
C_{+}\left(  \tilde{T}\right)  =C\left(  X=1_{+},\tilde{T}\right)  .
\label{eq34c}%
\end{equation}
Note that $C_{+}\left(  0\right)  =1,$ i.e. $c\left(  a,t_{a}\right)  =c_{s}.$

We observe that the problem for $X>1$ (i.e. $x>a$) decouples from that for
$X<1$ ($x<a$)$;$ we now solve these in turn.

\subsection{$X\geq1$}

First, we solve the problem for $X\geq1,\tilde{T}\geq0,$ corresponding to
$x\geq a,t\geq t_{a}.$ From Section \ref{pah}, the problem at hand is%
\begin{equation}
\frac{\partial C}{\partial\tilde{T}}=\frac{\partial^{2}C}{\partial X^{2}},
\label{f1}%
\end{equation}
subject to
\begin{align}
\frac{\partial C}{\partial X}  &  =0\quad\text{at }X=1,\label{f2}\\
C  &  \rightarrow0\quad\text{as }X\rightarrow\infty,\label{f3}\\
C  &  =C_{a}\left(  X\right)  \quad\text{at }\tilde{T}=0, \label{f4}%
\end{align}

Setting $\xi=X-1,$ we have%
\begin{equation}
\frac{\partial C}{\partial\tilde{T}}=\frac{\partial^{2}C}{\partial\xi^{2}},
\end{equation}
subject to
\begin{align}
\frac{\partial C}{\partial\xi}  &  =0\quad\text{at }\xi=0,\\
C  &  \rightarrow0\quad\text{as }\xi\rightarrow\infty,\\
C  &  =C_{a}\left(  \xi\right)  \quad\text{at }\tilde{T}=0,
\end{align}
where
\begin{equation}
C_{a}\left(  \xi\right)  =\frac{\text{erfc}\left(  \frac{a\left(
1+\xi\right)  -L_{d}}{2\sqrt{D_{w}t_{a}}}\right)  }{\text{erfc}\left(
-\frac{\theta}{2\sqrt{D_{w}}}\right)  }. \label{bigCa}%
\end{equation}

Thence, using Fourier transforms, we obtain%
\begin{equation}
C\left(  \xi,\tilde{T}\right)  =\frac{1}{2\sqrt{\pi\tilde{T}}}\int_{0}%
^{\infty}C_{a}\left(  X^{\prime}\right)  \left\{  \exp\left(  -\frac{\left(
\xi-X^{\prime}\right)  ^{2}}{4\tilde{T}}\right)  +\exp\left(  -\frac{\left(
\xi+X^{\prime}\right)  ^{2}}{4\tilde{T}}\right)  \right\}  \mathrm{d}%
X^{\prime}.
\end{equation}
Before we can tackle the second problem (i.e. the case $X<1),$ we shall
require $C_{+}\left(  \tilde{T}\right)  =C\left(  \xi=0,\tilde{T}\right)  $
for condition (\ref{eq34b})$,$ i.e.%
\begin{equation}
C_{+}\left(  \tilde{T}\right)  =\frac{1}{\sqrt{\pi\tilde{T}}}\int_{0}^{\infty
}C_{a}\left(  X^{\prime}\right)  \exp\left(  -\frac{X^{\prime^{2}}}{4\tilde
{T}}\right)  \mathrm{d}X^{\prime}. \label{2-53}%
\end{equation}
Putting $z=X^{\prime}/2\sqrt{\tilde{T}},$ we have%
\begin{align}
C_{+}\left(  \tilde{T}\right)   &  =\frac{2}{\sqrt{\pi}}\int_{0}^{\infty}%
C_{a}\left(  2z\sqrt{\tilde{T}}\right)  e^{-z^{2}}\mathrm{d}z\nonumber\\
&  =C_{a}\left(  0\right)  +\frac{2\sqrt{\tilde{T}}}{\sqrt{\pi}}%
\frac{\mathrm{d}C_{a}}{\mathrm{d}\xi}\left(  0\right)  +O\left(  \tilde
{T}\right)  , \label{Caa}%
\end{align}
where we have used a Taylor series expansion for $C_{a}$ about $z=0.$ Now, on
using (\ref{bigCa}) and recalling equation (\ref{theta}), we note that
$C_{a}\left(  0\right)  =1$ and that%
\[
\frac{\mathrm{d}C_{a}}{\mathrm{d}\xi}\left(  0\right)  =-\frac{a}{\sqrt{\pi
D_{w}t_{a}}}\frac{\text{exp}\left(  -\frac{\left(  a-L_{d}\right)  ^{2}%
}{4D_{w}t_{a}}\right)  }{\text{erfc}\left(  -\frac{\theta}{2\sqrt{D_{w}}%
}\right)  }.
\]
So, we have, for small $\tilde{T},$%
\begin{equation}
C_{+}\left(  \tilde{T}\right)  =1-\left\{  \frac{2a}{\pi\sqrt{D_{w}t_{a}}%
}\frac{\text{exp}\left(  -\frac{\left(  a-L_{d}\right)  ^{2}}{4D_{w}t_{a}%
}\right)  }{\text{erfc}\left(  -\frac{\theta}{2\sqrt{D_{w}}}\right)
}\right\}  \sqrt{\tilde{T}}+O\left(  \tilde{T}\right)  . \label{2.55}%
\end{equation}

However, to determine $C\left(  0,\tilde{T}\right)  $ for all $\tilde{T},$ we
need to revert to (\ref{2-53}) with $z=X^{\prime}/2\sqrt{\tilde{T}},$ which
gives%
\begin{equation}
C_{+}\left(  \tilde{T}\right)  =\frac{2}{\sqrt{\pi}\text{erfc}\left(
-\frac{\theta}{2\sqrt{D_{w}}}\right)  }\int_{0}^{\infty}\text{erfc}\left(
f\left(  z,\tilde{T}\right)  \right)  e^{-z^{2}}\mathrm{d}z,
\end{equation}
where%
\[
f\left(  z,\tilde{T}\right)  =\frac{a\left(  1+2z\sqrt{\tilde{T}}\right)
-L_{d}}{2\sqrt{D_{w}t_{a}}}.
\]
Differentiating with respect to $\tilde{T},$ we have%
\begin{equation}
\frac{\mathrm{d}C_{+}}{\mathrm{d}\tilde{T}}=-\frac{2a}{\pi\text{erfc}\left(
-\frac{\theta}{2\sqrt{D_{w}}}\right)  \sqrt{D_{w}t_{a}\tilde{T}}}\int
_{0}^{\infty}ze^{-\left(  z^{2}+f^{2}\left(  z,\tilde{T}\right)  \right)
}\mathrm{d}z. \label{diff}%
\end{equation}
Rearranging the argument in the exponential in (\ref{diff}), we have%
\[
z^{2}+\frac{\left(  2az\sqrt{\tilde{T}}+a-L_{d}\right)  ^{2}}{4D_{w}t_{a}%
}=\mathcal{A}\left(  \tilde{T}\right)  \left\{  \left(  z+\mathcal{B}\left(
\tilde{T}\right)  \right)  ^{2}+\mathcal{C}\left(  \tilde{T}\right)  \right\}
,
\]
where%
\begin{align}
\mathcal{A}\left(  \tilde{T}\right)   &  =1+\frac{a^{2}\tilde{T}}{D_{w}t_{a}%
},\\
\mathcal{B}\left(  \tilde{T}\right)   &  =\frac{\left[  \frac{\left(
a-L_{d}\right)  a\sqrt{\tilde{T}}}{D_{w}t_{a}}\right]  }{2\left(
1+\frac{a^{2}\tilde{T}}{D_{w}t_{a}}\right)  },\\
\mathcal{C}\left(  \tilde{T}\right)   &  =\frac{\frac{\left(  a-L_{d}\right)
^{2}}{4D_{w}t_{a}}}{\left(  1+\frac{a^{2}\tilde{T}}{D_{w}t_{a}}\right)
}-\frac{\left[  \frac{\left(  a-L_{d}\right)  a\sqrt{\tilde{T}}}{D_{w}t_{a}%
}\right]  ^{2}}{4\left(  1+\frac{a^{2}\tilde{T}}{D_{w}t_{a}}\right)  ^{2}};
\end{align}
it is now possible to write the integral in (\ref{diff}) in the form%
\begin{equation}
\int_{0}^{\infty}ze^{-\left\{  \mathcal{A}\left(  \tilde{T}\right)  \left[
\left(  z+\mathcal{B}\left(  \tilde{T}\right)  \right)  ^{2}+\mathcal{C}%
\left(  \tilde{T}\right)  \right]  \right\}  }\mathrm{d}z. \label{i}%
\end{equation}
Next, with $\zeta=z+\mathcal{B}\left(  \tilde{T}\right)  $ and later
$\xi=\mathcal{A}^{1/2}\left(  \tilde{T}\right)  \zeta,$ we have%
\begin{align}
&  \int_{0}^{\infty}ze^{-\mathcal{A}\left(  \tilde{T}\right)  \left[  \left(
z+\mathcal{B}\left(  \tilde{T}\right)  \right)  ^{2}+\mathcal{C}\left(
\tilde{T}\right)  \right]  }\mathrm{d}z\nonumber\\
&  =\frac{1}{2}e^{-\mathcal{A}\left(  \tilde{T}\right)  \mathcal{C}\left(
\tilde{T}\right)  }\left\{  \frac{e^{-\mathcal{A}\left(  \tilde{T}\right)
\mathcal{B}^{2}\left(  \tilde{T}\right)  }}{\mathcal{A}\left(  \tilde
{T}\right)  }-\frac{\pi^{1/2}\mathcal{B}\left(  \tilde{T}\right)
}{\mathcal{A}^{1/2}\left(  \tilde{T}\right)  }\text{erfc}\left(
\mathcal{A}^{1/2}\left(  \tilde{T}\right)  \mathcal{B}\left(  \tilde
{T}\right)  \right)  \right\}  .
\end{align}

Hence, we have the following first-order ordinary differential equation (ODE)
for $C_{+}\left(  \tilde{T}\right)  :$%
\begin{equation}
\frac{\mathrm{d}C_{+}}{\mathrm{d}\tilde{T}}=-\frac{ae^{-\mathcal{A}\left(
\tilde{T}\right)  \mathcal{C}\left(  \tilde{T}\right)  }}{\pi\text{erfc}%
\left(  -\frac{\theta}{2\sqrt{D_{w}}}\right)  \sqrt{D_{w}t_{a}}\sqrt{\tilde
{T}}}\left\{  \frac{e^{-\mathcal{A}\left(  \tilde{T}\right)  \mathcal{B}%
^{2}\left(  \tilde{T}\right)  }}{\mathcal{A}\left(  \tilde{T}\right)  }%
-\frac{\pi^{1/2}\mathcal{B}\left(  \tilde{T}\right)  }{\mathcal{A}%
^{1/2}\left(  \tilde{T}\right)  }\text{erfc}\left(  \mathcal{A}^{1/2}\left(
\tilde{T}\right)  \mathcal{B}\left(  \tilde{T}\right)  \right)  \right\}  ,
\label{ODEE}%
\end{equation}
subject to%
\begin{equation}
C_{+}=1\quad\text{at }\tilde{T}=0. \label{ODEE1}%
\end{equation}

Checking $\mathcal{A}\left(  \tilde{T}\right)  ,\mathcal{B}\left(  \tilde
{T}\right)  ,\mathcal{C}\left(  \tilde{T}\right)  $ in the limit as $\tilde
{T}\rightarrow0,$ we have%
\begin{equation}
\mathcal{A}\left(  0\right)  =1,\quad\mathcal{B}\left(  0\right)
=0,\quad\mathcal{C}\left(  0\right)  =\frac{\left(  a-L_{d}\right)  ^{2}%
}{4D_{w}t_{a}},
\end{equation}
so that%
\begin{equation}
\frac{\mathrm{d}C_{+}}{\mathrm{d}\tilde{T}}\sim\left(  -\frac{ae^{-\left(
a-L_{d}\right)  ^{2}/4D_{w}t_{a}}}{\pi\text{erfc}\left(  -\frac{\theta}%
{2\sqrt{D_{w}}}\right)  \sqrt{D_{w}t_{a}}}\right)  \frac{1}{\sqrt{\tilde{T}}}.
\end{equation}

\subsection{$X<1\label{prevsec}$}

For this region, we require to solve (\ref{eq34a})-(\ref{eq34c}). Note that,
from the solution for $X>1,$ we have already found in (\ref{2.55}) that, for
small $\tilde{T},$%
\begin{equation}
C_{+}\left(  \tilde{T}\right)  -1\sim\tilde{T}^{1/2}.
\end{equation}
Moreover, at $\tilde{T}=0,$ the region that we are solving in, i.e.
$0<\tilde{X}<\tilde{S}\left(  \tilde{T}\right)  ,$ has zero width, which
suggests that it may be appropriate to proceed in terms of similarity or
similarity-like variables. For this purpose, we set%
\begin{equation}
C-1=\tilde{T}^{1/2}F\left(  \eta,\tilde{T}\right)  ,\qquad\eta=\frac{\tilde
{X}}{\tilde{S}\left(  \tilde{T}\right)  }, \label{te3}%
\end{equation}
so that equation (\ref{eq34a}) becomes%
\begin{equation}
\frac{\tilde{S}^{2}\left(  \tilde{T}\right)  F}{2\tilde{T}}+\left(  \tilde
{S}^{2}\left(  \tilde{T}\right)  \frac{\partial F}{\partial\tilde{T}}%
-\tilde{S}\left(  \tilde{T}\right)  \frac{\mathrm{d}\tilde{S}}{\mathrm{d}%
\tilde{T}}\eta\frac{\partial F}{\partial\eta}\right)  =\frac{\partial^{2}%
F}{\partial\eta^{2}}, \label{w1}%
\end{equation}
subject to%
\begin{align}
1+\tilde{T}^{1/2}F  &  =C_{+}\left(  \tilde{T}\right)  \quad\text{at }%
\eta=0,\label{w2}\\
F  &  =0\quad\text{at }\eta=1,\label{w3}\\
-\frac{\partial F}{\partial\eta}  &  =\frac{\tilde{S}\left(  \tilde{T}\right)
}{\tilde{T}^{1/2}}\frac{\mathrm{d}\tilde{S}}{\mathrm{d}\tilde{T}}%
(1-\frac{c_{0}}{c_{s}})\quad\text{at }\eta=1. \label{w4}%
\end{align}
It is now required that (\ref{w1})-(\ref{w4}) behave in a self-consistent
manner as $\tilde{T}\rightarrow0;$ by this, we mean that we should obtain an
ODE, subject to the requisite number of boundary conditions.

To consider this systematically, start with equation (\ref{w1}) and suppose
that we try to retain as many terms on the left-hand side as possible as
$\tilde{T}\rightarrow0;$ this can be done if
\begin{equation}
\tilde{S}\left(  \tilde{T}\right)  \frac{\mathrm{d}\tilde{S}}{\mathrm{d}%
\tilde{T}}\sim1, \label{wrong}%
\end{equation}
which implies that $\tilde{S}\sim\tilde{T}^{1/2}$ and there is clearly a
sensible balance of leading order terms in (\ref{w1}) as $\tilde{T}%
\rightarrow0$. However, the right-hand side of equation (\ref{w4}) would
become unbounded as $\tilde{T}\rightarrow0,$ and hence (\ref{wrong}) does not
lead to overall self-consistency in this limit. Note also that if we try with%
\[
\tilde{S}\left(  \tilde{T}\right)  \frac{\mathrm{d}\tilde{S}}{\mathrm{d}%
\tilde{T}}\gg1,
\]
instead of (\ref{wrong}), then the left-hand side of (\ref{w1}) dominates the
right-hand side, and it will not be possible to satisfy all of the boundary
conditions as $\tilde{T}\rightarrow0.$ The only remaining possibility is if
\begin{equation}
\tilde{S}\left(  \tilde{T}\right)  \frac{\mathrm{d}\tilde{S}}{\mathrm{d}%
\tilde{T}}\ll1. \label{not}%
\end{equation}
To pin the behaviour down more precisely, we turn to (\ref{w4}), which
suggests that
\begin{equation}
\frac{\tilde{S}\left(  \tilde{T}\right)  }{\tilde{T}^{1/2}}\frac
{\mathrm{d}\tilde{S}}{\mathrm{d}\tilde{T}}\sim1, \label{right}%
\end{equation}
in order to balance with the term on the left-hand side. In this case, we
obtain $\tilde{S}\left(  \tilde{T}\right)  \sim\tilde{T}^{3/4},$ which ensures
a sensible leading-order balance in (\ref{w1}) and (\ref{w4}), noting also
that (\ref{not}) is fulfilled, since%
\[
\tilde{S}\left(  \tilde{T}\right)  \frac{\mathrm{d}\tilde{S}}{\mathrm{d}%
\tilde{T}}\sim\tilde{T}^{1/2}.
\]
Setting $\tilde{S}\left(  \tilde{T}\right)  =\lambda\tilde{T}^{3/4}+..,$ where
$\lambda$ is a positive constant to be determined, equation (\ref{w1})
becomes, in the limit as $\tilde{T}\rightarrow0,$%
\begin{equation}
\frac{\mathrm{d}^{2}F_{0}}{\mathrm{d}\eta^{2}}=0, \label{simode}%
\end{equation}
where
\begin{equation}
F_{0}\left(  \eta\right)  :=\lim_{\tilde{T}\rightarrow0}F\left(  \eta
,\tilde{T}\right)  .
\end{equation}
subject to
\begin{align}
F_{0}  &  =\mu\quad\text{at }\eta=0,\label{cb1}\\
F_{0}  &  =0\quad\text{at }\eta=1,\label{cb2}\\
-\frac{\mathrm{d}F_{0}}{\mathrm{d}\eta}  &  =\frac{3}{4}\lambda^{2}%
(1-\frac{c_{0}}{c_{s}})\quad\text{at }\eta=1, \label{cb3}%
\end{align}
where $\mu$ is a constant given by
\begin{equation}
\mu=\lim_{\tilde{T}\rightarrow0}\frac{\left(  C\right)  _{X=1}-1}{\tilde
{T}^{1/2}}. \label{muu}%
\end{equation}
Note that $\mu$ can be determined, and we will do so shortly, from the
solution for $X>1.$ Thus, solving (\ref{simode}) subject to (\ref{cb1}%
)-(\ref{cb3}) gives%
\begin{equation}
F_{0}\left(  \eta\right)  =\mu\left(  1-\eta\right)  ,
\end{equation}
with%
\begin{equation}
\mu=\frac{3}{4}\lambda^{2}(1-\frac{c_{0}}{c_{s}}),
\end{equation}
i.e.%
\begin{equation}
\lambda=\pm\left(  \frac{4\mu}{3(1-c_{0}/c_{s})}\right)  ^{1/2}. \label{lam}%
\end{equation}
Clearly, we need to take the positive sign to ensure that $\tilde{S}$
increases, i.e. $S$ decreases. Also, since $c_{0}>c_{s},$ it is clear that we
will need $\mu<0;$ we return to this point shortly.

Note also that it is possible to determine $\mu$ without solving
(\ref{f1})-(\ref{f4}). Near $X=1,$ we have
\begin{equation}
C_{a}=1+\left(  X-1\right)  \left(  \frac{\mathrm{d}C_{a}}{\mathrm{d}%
X}\right)  _{X=1}+...
\end{equation}
Now,%
\begin{equation}
C_{a}\left(  X\right)  =\frac{1+\text{erf}\left(  \frac{L_{d}-aX}{2\sqrt
{D_{w}t_{a}}}\right)  }{\text{erfc}\left(  -\frac{\theta}{2\sqrt{D_{w}}%
}\right)  },
\end{equation}
whence
\begin{equation}
\alpha:=\left(  \frac{\mathrm{d}C_{a}}{\mathrm{d}X}\right)  _{X=1}%
=-\frac{a\exp\left(  -\frac{\left(  L_{d}-a\right)  ^{2}}{4D_{w}t_{a}}\right)
}{\sqrt{\pi D_{w}t_{a}}\text{erfc}\left(  -\frac{\theta}{2\sqrt{D_{w}}%
}\right)  }. \label{alpha}%
\end{equation}
We consider the small and positive $X-1$ and small $\tilde{T}$ behaviour of
(\ref{f1})-(\ref{f4}) by setting $\xi=X-1,$ as after (\ref{f4}), and
\begin{equation}
C=1+\tilde{T}^{1/2}G\left(  \zeta,\tilde{T}\right)  ,\quad\zeta=\xi/\tilde
{T}^{1/2}. \label{use}%
\end{equation}
Equation (\ref{f1}) becomes%
\begin{equation}
\tilde{T}\frac{\partial G}{\partial\tilde{T}}+\frac{G}{2}-\frac{\zeta}{2}%
\frac{\partial G}{\partial\zeta}=\frac{\partial^{2}G}{\partial\zeta^{2}}.
\label{f5}%
\end{equation}
Now, in the limit as $\tilde{T}\rightarrow0,$ (\ref{f5}) becomes
\begin{equation}
\frac{G_{0}}{2}-\frac{\zeta}{2}\frac{\mathrm{d}G_{0}}{\mathrm{d}\zeta}%
=\frac{\mathrm{d}^{2}G_{0}}{\mathrm{d}\zeta^{2}}, \label{f6}%
\end{equation}
where
\begin{equation}
G_{0}\left(  \zeta\right)  :=\lim_{\tilde{T}\rightarrow0}G\left(  \zeta
,\tilde{T}\right)  .
\end{equation}
Equation (\ref{f6}) has the general solution%
\begin{equation}
G_{0}=K_{1}\zeta+K_{2}\left(  \pi\zeta\operatorname{erf}\left(  \frac{\zeta
}{2}\right)  +2\sqrt{\pi}\exp\left(  -\frac{\zeta^{2}}{4}\right)  \right)  ,
\end{equation}
where $K_{1}$ and $K_{2}$ are constants to be determined. Clearly, (\ref{f6})
must have two boundary conditions. One of these comes from (\ref{f2}), and is%
\begin{equation}
\frac{\mathrm{d}G_{0}}{\mathrm{d}\zeta}=0\quad\text{at }\zeta=0.
\end{equation}
The other comes from matching $G_{0}$ as $\zeta\rightarrow\infty$ to $C_{a}$
and is
\begin{equation}
\frac{\mathrm{d}G_{0}}{\mathrm{d}\zeta}\rightarrow\alpha\quad\text{as }%
\zeta\rightarrow\infty.
\end{equation}
Since
\begin{equation}
\frac{\mathrm{d}G_{0}}{\mathrm{d}\zeta}=K_{1}+K_{2}\pi\operatorname{erf}%
\left(  \frac{\zeta}{2}\right)  ,
\end{equation}
we quickly see that
\begin{equation}
K_{1}=0,\quad K_{2}=\frac{\alpha}{\pi},
\end{equation}
whence
\begin{equation}
G_{0}=\alpha\left(  \zeta\operatorname{erf}\left(  \frac{\zeta}{2}\right)
+\frac{2}{\sqrt{\pi}}\exp\left(  -\frac{\zeta^{2}}{4}\right)  \right)  ;
\end{equation}
ultimately, this leads to
\begin{equation}
\mu=G_{0}\left(  0\right)  =\frac{2\alpha}{\pi^{1/2}}. \label{mu}%
\end{equation}
Finally, recall from the discussion after equation (\ref{lam}) that we needed
$\mu<0.$ Now, equation (\ref{mu}) implies that we will need $\alpha<0;$ from
equation (\ref{alpha}), we see that this will clearly be the case.

\section{Results\label{sec3}}

\begin{figure}[ptb]
\centering\scalebox{0.6}{\includegraphics{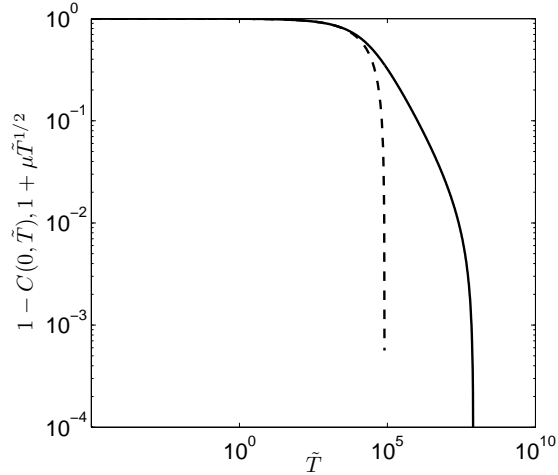}}
\caption{$1-C\left(  0,\tilde{T}\right)  $ (solid line) and $1+\mu\tilde
{T}^{1/2}$ (dashed line) vs. $\tilde{T}$ }%
\label{fig1}%
\end{figure}

The main numerical task is to solve equation (\ref{w1}), subject to
(\ref{w2})-(\ref{w4}); this constitutes a moving boundary problem for $F$ and
$\tilde{S}.$ However, (\ref{w2}) contains $C_{+}\left(  \tilde{T}\right)  ,$
which must itself be solved for numerically via the first-order ODE
(\ref{ODEE}), subject to (\ref{ODEE1}). To illustrate our ideas, we will vary
the value of $D_{e},$ so as to see the effect of $\delta,$and select the
following parameters from \cite{Vo:2018}: $L_{d}=10^{-5}$ m, $a=0.2L_{d},$
$D_{w}=5\times10^{-11}$ m$^{2}$s$^{-1},$ $c_{0}/c_{s}=50.$

However, before presenting the results, we note first that we are ultimately
interested in determining the time at which the front reaches $x=0;$ this
corresponds to the time at which $\tilde{S}=1/\delta^{1/2}.$ Whilst this will,
of course, depend on the value of $\delta,$ we observe that $c_{s}-c\left(
a,t\right)  ,$ and hence $1-C\left(  \tilde{X}=0,\tilde{T}\right)  ,$ i.e.
$1-C_{+}\left(  \tilde{T}\right)  $, will be independent of $\delta;$ this is
evident since there is no $\delta$ in either equation (\ref{ODEE}) or
(\ref{ODEE1}). Thus, it makes sense to look at $1-C\left(  0,\tilde{T}\right)
$ vs. $\tilde{T},$ ahead of considering the solutions for $\tilde{S}$ and
$C\left(  X,\tilde{T}\right)  .$ Thus, Fig. \ref{fig1} shows a log-log plot
for $1-C\left(  0,\tilde{T}\right)  $ vs. $\tilde{T},$ as well $1+\mu\tilde
{T}^{1/2}$ vs. $\tilde{T};$ the second of these is the small-time
approximation for $1-C\left(  0,\tilde{T}\right)  $ derived in Section
\ref{prevsec} and makes use of the form for $C$ in (\ref{use}) and (\ref{mu}).
We see that this approximation works quite well until $\tilde{T}\sim10^{4},$
after which the two curves diverge.

\begin{figure}[ptb]
\centering\scalebox{0.6}{\includegraphics{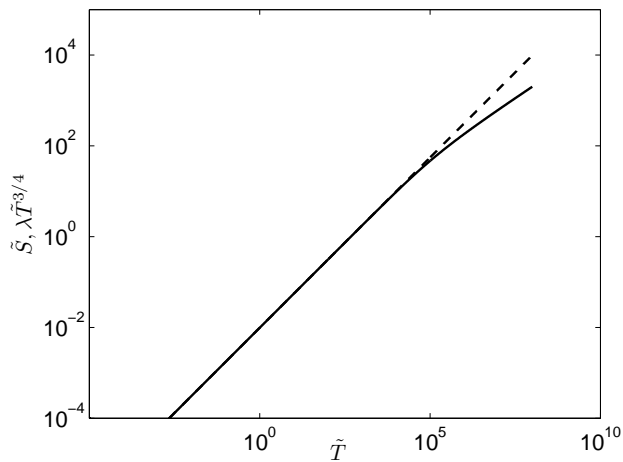}}
\caption{$\tilde{S}$ (solid line, computed) and $\lambda\tilde{T}^{3/4}$
(dashed line) vs$.$ $\tilde{T}.$ Note that the computation has been stopped at
$\tilde{T}=10^{8};$ at this stage $\tilde{S}$ $\approx9823,$ which implies
that $\delta\approx10^{-8}.$ In more detail, with $\bar{S}=1/\delta^{1/2},$ we
have $\delta=1/9823^{2}=$ 1.0364$\times10^{-8}.$}%
\label{fig2}%
\end{figure}

Next, Fig. \ref{fig2} shows $\tilde{S}$ vs$.$ $\tilde{T},$ as well as
$\lambda\tilde{T}^{3/4}$ vs$.$ $\tilde{T};$ the latter of these is also from
the small-time approximation, as indicated between equations (\ref{right}) and
(\ref{simode}). Whilst this result does not depend on $\delta$ either$,\ $we
have stopped the computation when $\tilde{S}$ reaches $O\left(  10^{4}\right)
$, with a view to exploring the results when $\delta\geq10^{-8};$ this covers
the range in $\delta$ considered in \cite{Vo:2018}. Here also, we see that the
two curves follow each other until $\tilde{T}\sim10^{4},$ at which point
$\tilde{S}\sim10^{2}$. This would mean that, for 10$^{-4}\leq\delta\ll1,$ a
preliminary estimate for $\tilde{T}$ of when $\tilde{S}=1/\delta^{1/2},$ which
we denote by $\tilde{T}_{stop},$ would be given by
\begin{equation}
\lambda\tilde{T}_{stop}^{3/4}\approx\frac{1}{\delta^{1/2}}, \label{tstop}%
\end{equation}
giving $\tilde{T}_{stop}\approx\left(  \lambda\delta^{1/2}\right)  ^{-4/3}.$
In actual time, this amounts to%
\begin{equation}
t_{stop}:=a^{2}\delta\tilde{T}_{stop}/D_{e}\left(  =a^{2}\tilde{T}%
_{stop}/D_{w}\right)  , \label{tstop2}%
\end{equation}
where $t_{stop}$ is the time taken for the front to move from $X=1$ to $X=0,$
i.e. $x=a$ to $x=0.$

However, the values for $\delta$ used in \cite{Vo:2018} lie outside of this
range - they are smaller - and any attempt to use equation (\ref{tstop}) can
thus be expected to underestimate the value of $t_{stop}.$ Instead, in Table
\ref{tab1}, we compare the values of $t_{stop}$ as given by the solid line in
Fig. \ref{fig2}, which were obtained from the solution of (\ref{w1}%
)-(\ref{w4}), and as estimated from Fig. 3 in \cite{Vo:2018}, for different
values of $\delta$. As can be seen, the qualitative and quantitative agreement
is very good.

\begin{table}[ptb]
\begin{center}%
\begin{tabular}
[c]{|c|c|c|}\hline
$\delta$ & \multicolumn{2}{|c|}{$t_{stop}\left[  \text{days}\right]  $%
}\\\cline{2-3}
& Fig. \ref{fig2} & \cite{Vo:2018}\\\hline
$5\times10^{-7}$ & $\sim$46.4 & $\sim$46.5\\\hline
$10^{-6}$ & $\sim$23.8 & $\sim$23\\\hline
$5\times10^{-6}$ & $\sim$4.97 & $\sim$5\\\hline
$10^{-5}$ & $\sim$2.6 & $\sim$2.5\\\hline
\end{tabular}
\end{center}
\caption{$t_{stop},$ as calculated in two different ways for four values of
$\delta.$ }%
\label{tab1}%
\end{table}

\begin{figure}[ptb]
\centering\scalebox{0.6}{\includegraphics{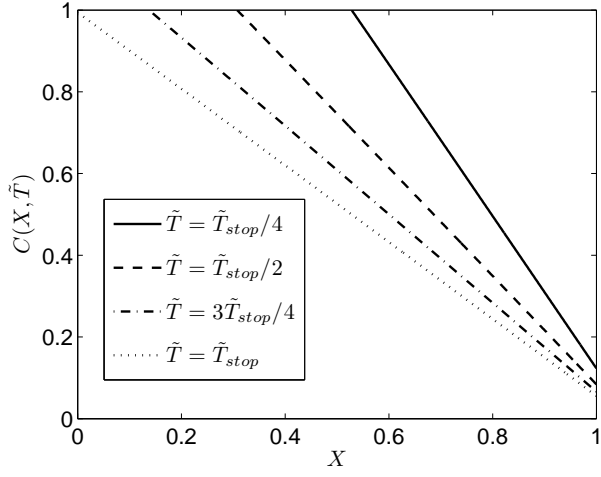}}
\caption{$C$ vs$.$ $X$ for four different values of $\tilde{T}$ for
$\delta=10^{-5}$ . \ $\tilde{T}_{stop}$ corresponds to $t_{stop}=$2.6 days.}%
\label{fig4}%
\end{figure}

\begin{figure}[ptb]
\centering\scalebox{0.6}{\includegraphics{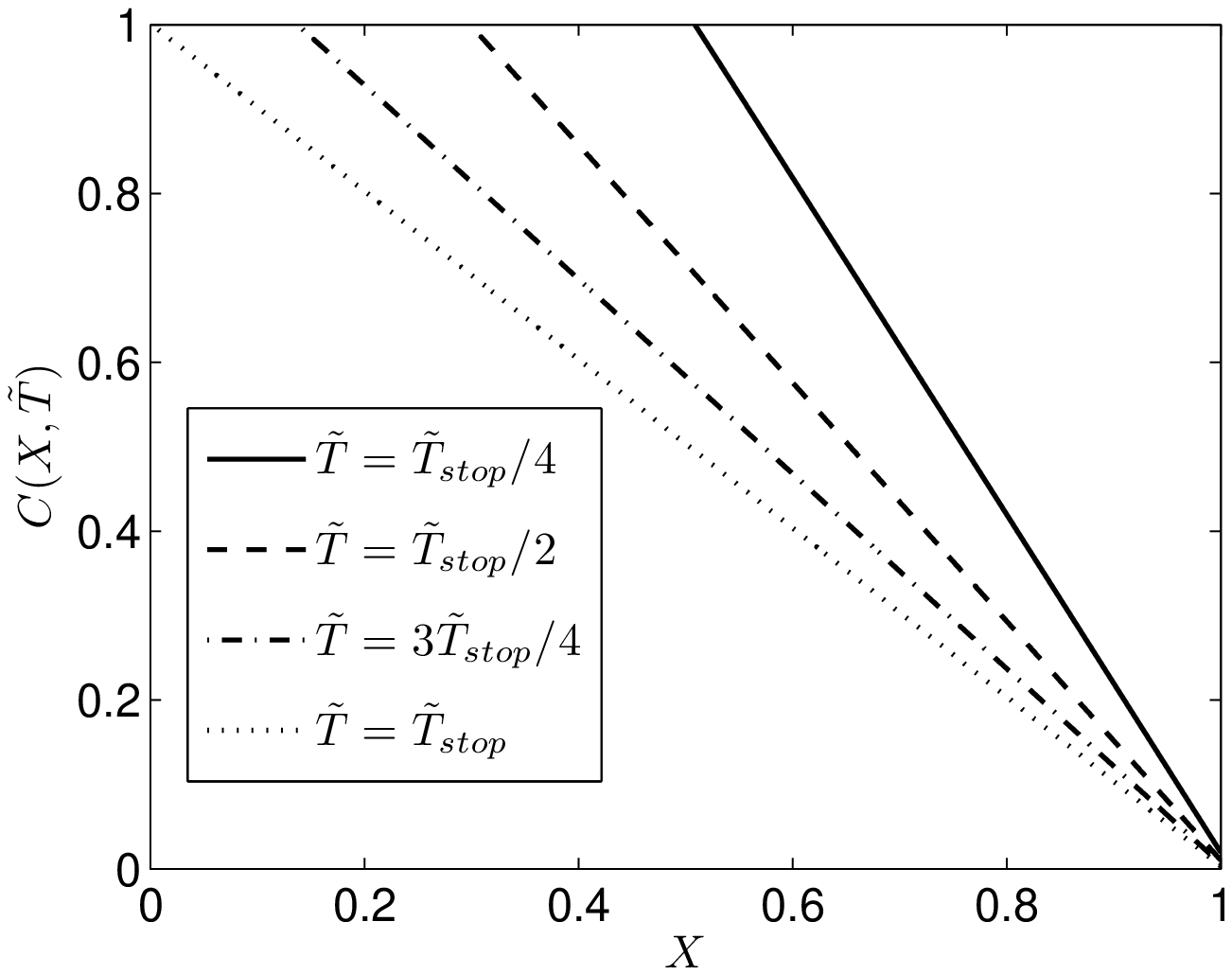}}
\caption{$C$ vs$.$ $X$ for four different values of $\tilde{T}$ for
$\delta=5\times10^{-7}$ . \ $\tilde{T}_{stop}$ corresponds to $t_{stop}=$46.4
days.}%
\label{fig5}%
\end{figure}

An interesting observation now arises: if $D_{w},$ $L_{d}/a$ and $c_{0}/c_{s}$
are fixed, only one computation, i.e. the one that was already carried out to
determine the profile for $\tilde{S}$ for $\tilde{T}$ as great as 10$^{8}$
already and which generated the results for Fig. \ref{fig2}, is required to
find the solution for $C\left(  X,\tilde{T}\right)  ,$ which comes from the
solution for $F$ via equation (\ref{te3}), for any value of $\delta!$ This is
as opposed to having to carry out a new computation on each occasion that
$D_{e},$ and hence $\delta,$ is changed, as was done in \cite{Vo:2018}. To see
this, we show in Figs. \ref{fig4} and \ref{fig5} $C$ as a function of $X$ for
$\tilde{S}\left(  \tilde{T}\right)  \leq X\leq1$ for four different values of
$\tilde{T}$ for $\delta=10^{-5}$ and 5$\times$10$^{-7},$ respectively; note
that, in these figures, the concentration profile at $\tilde{T}=0,$
corresponding to $t=t_{a},$ consists of a point that is located at $C=1$ and
$X=1$ but which then become a curve - a line, as it turns out - that moves
down and to the left with time. In both figures, $X$ is related to the
independent variables of the domain in which the computations were carried
out, $\eta$ and $\tilde{T},$ by%
\[
X=1-\delta^{1/2}\eta\tilde{S}\left(  \tilde{T}\right)  ,
\]
as can be seen by tracking back through the substitutions in equations
(\ref{three-eight}), (\ref{te2}) and (\ref{te3}).

\section*{Acknowledgment}

The first author would like to acknowledge the award of a Sir David Anderson
Bequest from the University of Strathclyde.


\bibliographystyle{abbrv}
\bibliography{stentbib}




\end{document}